\documentclass[12pt]{iopart}
\usepackage[dvips]{epsfig}
\usepackage{iopams}
\usepackage{appendix, color}

\begin{document}

\title[Generation of coherent states of photon-added type via pathway of eigenfunctions]
{Generation of coherent states of photon-added type via pathway of eigenfunctions}
\author{K. G\'{o}rska$^{a, b}$, K. A. Penson$^{b}$
and G. H. E. Duchamp$^{c}$\vspace{2mm}}

\address
{$^a$ Nicolaus Copernicus University, Institute of Physics, ul. Grudzi\c{a}dzka 5/7,\\
PL 87-100 Toru\'{n}, Poland\vspace{2mm}}

\address
{$^b$ Laboratoire de Physique Th\'eorique de la Mati\`{e}re Condens\'{e}e,\\
Universit\'e Pierre et Marie Curie, CNRS UMR 7600\\
Tour 24 - 2i\`{e}me \'et., 4 pl. Jussieu, F 75252 Paris Cedex 05, France\vspace{2mm}}

\address
{$^c$ Universit\'e Paris XIII, LIPN, Institut Galil\'{e}e, CNRS UMR 7030, \\
99 Av. J.-B. Clement, F 93430 Villetaneuse, France \vspace{2mm}}

\eads{\linebreak \mailto{dede@fizyka.umk.pl}, 
\mailto{penson@lptl.jussieu.fr}, 
\mailto{ghed@lipn-univ.paris13.fr}}


\begin{abstract}
\\
We obtain and investigate the regular eigenfunctions of simple differential operators $x^r\,  d^{r+1}/dx^{r+1}$, $r=1, 2, \ldots$ with the eigenvalues equal to one. With the help of these eigenfunctions we construct a non-unitary analogue of boson displacement operator which will be acting on the vacuum. In this way we generate collective quantum states of the Fock space which are normalized and equipped with the resolution of unity with the positive weight functions that we obtain explicitly. These states are thus coherent states in the sense of Klauder. They span the truncated Fock space without first $r$ lowest-lying basis states: $|0\rangle, |1\rangle, \ldots, |r-1\rangle$. These states are squeezed, are sub-Poissonian in nature and are reminiscent of photon-added states at Agarwal et al.
\end{abstract}

\maketitle

\section{Introduction}

We are concerned in this work with a procedure of generating collective quantum states spanning a prescribed infinite subset of the Fock space $\{|n\rangle\}_{n=0}^{\infty}$. The basis states $|n\rangle$ are orthonormal, $\langle n| n'\rangle = \delta_{n, n'}$ and are eigenstates of the boson number operator $\hat{n} = a^{\dag}a$, $\hat{n} |n\rangle = n|n\rangle$, where $a^{\dag}$ and $a$ are the boson creation and annihilation operators satisfying $[a, a^{\dag}] = 1$ and $a|0\rangle = 0$.

Our objective is to generate in a systematic way families of states sharing their main properties with the so-called standard coherent states (CS). These states are the CS in the sense of Klauder \cite{JRKlauder85}. It means that they are: a) normalized, b) continuous with their label (generally a complex number $z$) and c) they admit the resolution of unity with a positive weight function.

The central tools in the domain of standard CS is the unitary displacement operator $\mathcal{D}(z)=\exp(z a^{\dag} - z^{*} a)$ whose action on the Fock vacuum $|0\rangle$ generates the normalized standard CS denoted by $|z\rangle$ \cite{LEBallantine98}
\begin{eqnarray}\label{eqI1}
|z\rangle &=& \mathcal{D}(z)|0\rangle \\[0.7\baselineskip]
&=& \exp(-|z|^2/2) \, e^{z\, a^{\dag}}\, |0\rangle \label{eqI2}.
\end{eqnarray}
Note that the state $|z\rangle$ is the eigenstate of the annihilation operator $a$:
\begin{equation}\label{eqI2a}
a\,|z\rangle\,=\,z\,|z\rangle.
\end{equation}
We expose now a series of arguments leading to a novel approach to construct families of CS. This approach lends itself to many generalizations. 

First of all let us consider quantum states similar to (\ref{eqI1}) and (\ref{eqI2}) but created with higher powers of boson creation operator. They may be generated by powers of $a^{\dag}$ appearing in the exponential, via
\begin{equation}\label{eqI3}
|z, p\rangle \sim e^{z (a^{\dag})^p} |0\rangle.
\end{equation}
However, the convergence arguments forbid the creation of states $|z, p\rangle$ for $p > 2$, as they cannot be normalized, contradicting the requirement a) above. Thus it is impossible to get the higher order CS in this manner \cite{RAFisher84}. A possible way out from this dilemma is to attempt to use functions other that the exponential assuring that the states are normalizable. 

From now on, let us employ the formal operational equivalence \cite{KAPenson09}
\begin{equation}\label{eqI2b}
\left[\frac{d}{d x}, x\right] = 1 \Leftrightarrow \left[a, a^{\dag}\right] = 1,
\end{equation}
which permits one to relate $d/dx$ and $x$ respectively to the annihilation $a$ and creation $a^{\dag}$ operators. In this sense the displacement operator acting on vacuum, see Eq.(\ref{eqI2}) is equivalent to $e^{z x}$, which in turn  is the eigenfunction of the operator $d/dx$ with the eigenvalue $z$, reminiscent of Eq.(\ref{eqI2a}). Following this initial idea we will seek eigenfunctions of operators more general than $d/d x$ and shall attempt to construct CS associated with them.

In this paper we introduce a new generalization of the boson displacement  operator enabling to construct in the systematic way the coherent states of photon-added type \cite{GSAgarwal91}.   

The plan of the paper is the following: in Sec. \ref{SecEigen} we consider the simple differential operators $x^r d^{r+1}/d x^{r+1}$, $r=1, 2, ...$ and find their eigenfunctions. Then, in Sec. \ref{SecGenCS} with the help of these eigenfunctions and by using the relations (\ref{eqI2b}),  we construct a new non-unitary analogue of displacement operator. The action of so obtained displacement operator on $|0\rangle$ yields quantum states satisfying the requirements a), b) and c) above. We demonstrate that these CS bear analogy to the so-called photon-added states \cite{GSAgarwal91}. In Sec. \ref{SecEigenProper} we show that these CS satisfy certain eigenequation analogous to Eq.(\ref{eqI2a}) above. Further we define physical and statistical properties of these states, like the Mandel parameter, the metric factor and the squeezing. The resolution of unity and the associated Stieltjes moment problem are considered in Sec. \ref{SecRes}. In Sec. \ref{SecFig} we present many examples of properties of CS which were introduced in Sec. \ref{SecEigenProper}, calculated for various values of parameter $r$. The last Sec. \ref{SecDis} is devoted to discussion and conclusions. In Appendix we discuss the nature of the Stieltjes moment problem of Sec. \ref{SecRes}. We demonstrate the non-unique character of solutions of the resolution of unity obtained in Sec. \ref{SecRes}.

\section{Eigenfunctions of differential operators}
\label{SecEigen}

The starting point of our subsequent construction of quantum states is the determination of the eigenstates of differential operators in the form $x^{r} d^{r+1}$, $r = 1, 2, \ldots$, where $d = \frac{d}{dx}$, i.e. finding the eigenfunctions (with the eigenvalues equal to one) denoted by $E_{q}(r, x)$ satisfying
\begin{equation}\label{eq1}
x^r d^{r+1} E_{q}(r, x) = E_{q}(r, x), \quad r = 1, 2, \ldots, \quad q = 1, \ldots, r+1.
\end{equation}
The Eq.(\ref{eq1}) is an ordinary differential equation of order $r + 1$ and it possesses $r + 1$ independent solutions labelled by $q$, but only one of them is Taylor expandable at $x=0$, see \cite{Burbaki03}, and  precisely this one, denoted by $E(r, x)$, is of interest for our applications.

The implementation of Eq.(\ref{eq1})  via Frobenius recursive method \cite{GArfken} gives the explicit form:
\begin{equation}\label{eq3} 
E(r, x) = x^r \cdot\,_{0}F_{r}([\,\,\,], [2, 3, \ldots, r+1], x),
\end{equation}
which satisfies the following $r+1$ "initial" conditions at $x=0$:
\begin{equation}\label{eq2} 
E(r, 0) = 0, \,\, \frac{d^p E(r, x)}{d x^p}\left|\right._{x=0} = 0, \, p = 1, \ldots, r-1, \,\, \frac{d^r E(r, x)}{d x^r}\left|\right._{x=0} = r!\,\, . \quad
\end{equation}
In Eq.(\ref{eq3}) $\,_{p}F_{q}(\ldots)$ is a generalized hypergeometric function \cite{przypis1, APPrudniko98v}. Note that the Taylor expansion of $E(r, x)$ at $x = 0$ starts with $x^r$. Only the lowest order case $r = 1$ can be written  down in terms of standard functions:
\begin{equation}\label{eq4}
E(1, x) = \sqrt{x}\, I_{1}(2\sqrt{x}),
\end{equation}
where $I_{1}(y)$ is the modified Bessel function of the first kind. For $r > 1$ the functions $E(r, x)$ are related to the so-called hyper-Bessel functions \cite{OIMarichev83, YBCheikh98, VSKirykova93}. (Note in passing that the irregular solutions of Eq.(\ref{eq1}), $\tilde{E}_{q}(r, x)$ with $\lim_{x\to 0}\tilde{E}_{q}(r, x) = \infty$, are explicitly known only in the case $r = 1$: $\tilde{E}_{1}(1, x) = \sqrt{x} K_{1}(2 \sqrt{x})$, where $K_{1}(y)$ is the modified Bessel function of the second kind. These solutions do not enter our considerations).

\section{Generation of coherent states $|z\rangle_{r}$}
\label{SecGenCS}

We introduce the boson creation and annihilation operators $a^{\dag}$ and $a$, obeying the commutation relation $[a, a^{\dag}] = 1$ and we associate with them the Fock space of normalized and orthogonal vectors $|n\rangle$, $\langle n | n' \rangle = \delta_{n,\, n'}$, $\,\,n,\,n' = 0, 1, \ldots,$  satisfying the usual relations $a^{\dag} |n\rangle = \sqrt{n+1} |n+1\rangle$ and $a^{\dag} a|n\rangle = \hat{n}|n\rangle = n|n\rangle$, which is the eigenequation of the Hamiltonian of the harmonic oscillator.

In order to incorporate the standard CS into a framework and notation that we develop later in this paragraph we note that the exponential $\exp(z\, x)$ can be written as $_{0}F_{0}([\,\,\,], [\,\,\,], z\, x )$ and consequently we rewrite $\mathcal{D}(z)$ acting on $|0\rangle$ via Eq.(\ref{eqI1}) as:
\begin{equation}\label{eqG1}
\frac{_{0}F_{0}([\,\,\,], [\,\,\,], z\, a^{\dag})}{_{0}F_{0}([\,\,\,], [\,\,\,], |z|^2/2)} \, |0\rangle.
\end{equation}
We extend this idea to higher order operators appearing in Eqs.(\ref{eq1}) and (\ref{eq3}) by defining the analogue of the displacement operator relevant to $x^r d^{\,r+1}$ as \textit{proportional} to $E(r, z a^{\dag})$ given by Eq.(\ref{eq3}). 

To do so, for a complex number $z$ we introduce now a continuously parametrized family of normalized quantum states. They are defined by using a non-unitary analogue $\widetilde{\mathcal{D}_{r}}(z)$ of the displacement operator which is proportional to $E(r, z a^{\dag})$, acting on the vacuum $|0\rangle$:
\begin{eqnarray}\label{eq5a}
|z\rangle_{r} &\equiv& \widetilde{\mathcal{D}_{r}}(z) |0\rangle_{r} = \mathcal{N}_{r}^{-1/2}(|z|^2) \frac{E(r, z a^{\dag})}{z^r \, b(r)} \, |0\rangle \\[0.7\baselineskip] \label{eq5b}
&=& \frac{\mathcal{N}_{r}^{-1/2}(|z|^2)}{b(r)} \, (a^{\dag})^r \,_{0}F_{r}([\,\,\,], [2, 3, \ldots, r+1], z a^{\dag}) |0\rangle,
\end{eqnarray}
where $b(r)$ is a numerical factor equal to $\prod_{k = 0}^{r} k!$ and $\mathcal{N}_{r}(|z|^2)$ is the normalization function obtained from the condition $\,_{r}\langle z|z\rangle_{r} = 1$:
\begin{equation}\label{eq6}
\mathcal{N}_{r}(x) = \left(r!\,\prod_{k = 0}^{r - 1} (k!)^2\right)^{-1}\ _{0}F_{2 r} ([\,\,\,], [1, \underbrace{2, 2, \ldots, r, r}_{2r-2 \, {entries}}, r+1], x),\,\, x \geq 0
\end{equation}
which is a perfectly converging function for all values of $x$ \cite{APPrudniko98v}. Formulae (\ref{eq5a}) and (\ref{eq5b}) are key equations of our method. 

We identify the numerical factor in front of $\,_{0}F_{2r}$ in Eq.(\ref{eq6}) as $\rho_{r}^{-1}(0)$ for reasons which will become clear later, see Eq.(\ref{eq15}) below. Note that in Eq.(\ref{eq5a}) $\widetilde{\mathcal{D}_{r}}(0)=1$. The overlapping factor $\,_{r}\langle z | z' \rangle_{r}$ is equal to $\mathcal{N}_{r} (z^{\star}\cdot z')$, which is everywhere non-vanishing, except for a set (of measure zero) of \textit{zeros} of $\mathcal{N}_{r}(y)$ in the complex plane $y$. Some informations about the localizations of zeros of the function $\mathcal{N}_{r}(y)$ can be obtained from the Hurwitz criterion \cite{SGKrantz99} but this is beyond the scope of this work. 

The expansion of $|z\rangle_{r}$ in terms of basic states $|n\rangle$ reads from Eq.(\ref{eq5b}):
\begin{equation}\label{eq7}
|z\rangle_{r} = \mathcal{N}_{r}^{-1/2}(|z|^2) \, \sum_{n=0}^{\infty} \frac{z^n}{n! (n+1)! \ldots (n+r-1)! \sqrt{(n+r)!}} |n+r\rangle,
\end{equation}
from which we deduce that $|z\rangle_{r}$ is spanned on the Fock space \textit{without} first $r$ lowest-lying states: $|0\rangle, |1\rangle, \ldots, |r-1\rangle$. The reason for that is explained above, see the remark after Eq.(\ref{eq2}). As such, the $|z\rangle_{r}$'s are reminiscent of but not directly  related to the so-called photon-added coherent states, introduced in \cite{GSAgarwal91} and later studied in \cite{SSivakumar99, JMSixdeniers01, MDaoud02, MNHounkonnou091, DPopov02}.

For any hermitian Hamiltonian $\mathcal{H}$, in particular the harmonic oscillator Hamiltonian $H = a^{\dag}a + 1/2$, the time evolution of state $|z\rangle_{r} \equiv |z, t=0\rangle_{r}$ is given by $|z, t\rangle_{r} = e^{-i\mathcal{H} t}|z, 0\rangle_{r}$, which immediately implies that $\mathcal{N}_{r}(|z|^2)$ is constant in time.

\section{Eigenproperties of coherent states $|z\rangle_{r}$ and some expectation values}
\label{SecEigenProper}

In full analogy with many known varieties of coherent states \cite{MNHounkonnou091, MNHounkonnou092, AISolomon94, CQuesne02} the state $|z\rangle_{r}$ turns out to be an eigenstate of a certain generalized boson annihilation operator. In fact we show that this operator is equal to $(a^{\dag})^r\, a^{r+1}$:
\begin{eqnarray}\label{eq8} 
(a^{\dag})^r a^r \cdot a |z\rangle_{r} &=&  z |z\rangle_{r} \\[0.5\baselineskip]
&=& a \left[\sum_{k = 1}^{r} \sigma(r, k) (n-1)^k \right] |z\rangle_{r} = a \prod_{k = 1}^{r} (n - k) |z\rangle_{r} \label{eq9} \\[0.5\baselineskip] 
&=&  a \left[\sum_{k = 1}^{r+1} |\sigma(r+1, k)| (n-r-1)^{k-1} \right] |z\rangle_{r}, \label{eq10}
\end{eqnarray}  
where $\sigma(r, k)$ are the conventional Stirling numbers of the first kind \cite{MAbramowitz72}, and $n = a^{\dag} a$. In order to prove the eigenproperty of Eq.(\ref{eq8}) it suffices to observe that in the expansion of Eq.(\ref{eq7}) the use of relation 
\begin{equation}\label{eq10a}
(a^\dag)^r a^{r+1} |n+ r\rangle = \frac{(n-1+r)!}{(n-1)!} \, \sqrt{n+r} \, |n+r-1\rangle,
\end{equation}
and a standard change of summation index immediately gives the required result. Furthermore, the Eq.(\ref{eq9}) is a consequence of the generating function for $\sigma(r, k)$ \cite{MAbramowitz72} whereas its alternative form in Eq.(\ref{eq10}) comes from a slight modification of Eq.(11) in Ref. \cite{KAPenson09}.

Since Eqs.(\ref{eq9}) and (\ref{eq10}) imply that 
\begin{equation}\label{eq12}
a f_{r}(n) \,|z\rangle_{r} = z\, |z\rangle_{r},
\end{equation}
with $f_{r}(x) = \sum_{k = 1}^{r} \sigma(r, k) (x - 1)^k$, which is a simple function of $x$, the states $|z\rangle_{r}$ fall into the category of non-linear coherent states, frequently discussed in the literature \cite{ AISolomon94, VIManko97, SSivakumar98, BRoy00, MNHounkonnou07} in different contexts.

The knowledge of Fock-space expansion of $|z\rangle_{r}$ of Eq.(\ref{eq7}) allows one to obtain explicitly the expectation values of many operators involving $a$'s and $a^{\dag}$'s. Below we quote some of them:
\begin{equation}\label{eq13}
\,_{r}\langle z|(a^{\dag})^p a^p|z\rangle_{r} = \frac{x^{p-r}}{\mathcal{N}_{r}(x)}\, \frac{d^{p}}{d x^{p}} \left[x^r \mathcal{N}_{r}(x)\right], \quad x = |z|^2, \quad p = 0, 1, \ldots . 
\end{equation}
and 
\begin{equation}\label{eq14}
\,_{r}\langle z|(a^{\dag})^p a^s|z\rangle_{r} = \frac{(z^{\star})^p z^s }{\mathcal{N}_{r}(|z|^2)} \, \sum_{n=0}^{\infty} \left[\frac{(n+r)! (n+r+p-s)!}{\rho_{r}(n) \rho_{r}(n+p-s)}\right]^{1/2} \frac{|z|^{2 n}}{(n+r-s)!},
\end{equation}
where $p, s = 0, 1, \ldots$ and
\begin{equation}\label{eq15}
\rho_{r}(n) = \left[\prod_{k=0}^{r-1}(n+k)!\right]^2 (n+r)!, \quad \rho_{r}(0) = \left(\prod_{k=0}^{r-1}k!\right)^{2} r!, \,\, r=1, 2, \ldots.
\end{equation}  
Armed with Eqs.(\ref{eq13}) and (\ref{eq14}) the following physical and statistical characteristics of the state $|z\rangle_{r}$ can be readily calculated:

\noindent
- The probability $P_{r}(k, x)$ of finding the vector $|k\rangle$ in the state 
$|z\rangle_{r}$, $k\geq r$:
\begin{equation}\label{eq16}
P_{r}(k, x) = \frac{1}{\mathcal{N}_{r}(x)} \, \frac{x^{k-r}}{\rho_{r}(k-r)}.
\end{equation}

\noindent
- The average number $\bar{n}_{r}(x)$ of bosons in the state $|z\rangle_{r}$:
\begin{equation}\label{eq16a}
\bar{n}_{r}(x) = \,_{r}\langle z| n| z\rangle_{r} = \sum_{k=r}^{\infty} k\, P_{r}(k, x).
\end{equation}
More generally $\overline{n^{\,p}}_{r}(x) = \,_{r}\langle z| n^p | z\rangle_{r}$, $p=1, 2, \ldots\,\,$.

\noindent
- The Mandel parameter $Q_{M, r}(x)$, which yields information about the deviation of the probability distribution $P_{r}(k, x)$ from the Poisson distribution, is defined as (see Ref. \cite{LMandel95}):
\begin{equation}\label{eq17}
Q_{M, r}(x) = \frac{\overline{n^2}_{r}(x) - [\bar{n}_{r}(x)]^{2}}{\bar{n}_{r}(x)} - 1 \, = \, x \left[\frac{\left(x^r \mathcal{N}_{r}(x)\right)''}{\left(x^r \mathcal{N}_{r}(x)\right)'} - \frac{\left(x^r \mathcal{N}_{r}(x)\right)'}{x^r \mathcal{N}_{r}(x)}\right].
\end{equation}  
As it is the well-known \cite{LMandel95} in the Poissonian case, relevant for standard CS, we have $Q_{M, r}(x) = 0$ while for $Q_{M, r}(x) < 0$ (resp. $Q_{M, r}(x) >0$) we say that the distribution is sub-Poissonian (resp. super-Poissonian).

\noindent
- The metric factor $\omega_{r}(x)$:
\begin{equation}\label{eq17a}
\omega_{r}(x) = \left[x \frac{\mathcal{N}'_{r}(x)}{\mathcal{N}_{r}(x)}\right]' .
\end{equation}
It describes the geometric features of coherent states: $\omega_{r}(x) = 1$ for standard CS and a deviation from it measures the non-standard behaviour of states in question \cite{JRKlauder01}.

\noindent
- Squeezing in both $X$ and $P$, where $X$- and $P$-quadratures are given in terms of $a$ and $a^{\dag}$ by usual formulae
\begin{equation}\label{eq17i}
X = (a + a^{\dag})/\sqrt{2}, \quad P = -i (a - a^{\dag})/\sqrt{2}.
\end{equation}
The uncertainties in $X$ and $P$ are given in terms of standard formulae \cite{SSivakumar99}
\begin{eqnarray}\nonumber
(\Delta X)^2 &=& \overline{X^2} - \bar{X}^2 \\[0.5\baselineskip] \label{eq17b}
&=& \frac{1}{2}\left(1 + 2\langle a^{\dag}a\rangle + \langle a^2\rangle + \langle {a^{\dag}}^2\rangle - \langle a\rangle^2 - \langle a^{\dag}\rangle^2 - 2 \langle a\rangle\langle a^{\dag}\rangle\right),
\end{eqnarray}
\begin{eqnarray}\nonumber
(\Delta P)^2 &=& \overline{P^2} - \bar{P}^2 \\[0.5\baselineskip]
&=& \frac{1}{2}\left(1 + 2\langle a^{\dag}a\rangle - \langle a^2\rangle - \langle {a^{\dag}}^2\rangle + \langle a\rangle^2 + \langle a^{\dag}\rangle^2 - 2 \langle a\rangle\langle a^{\dag}\rangle\right), \label{eq17c}
\end{eqnarray}
where in Eqs.(\ref{eq17b}) and (\ref{eq17c}) all the averages $\langle \cdots\rangle$ are understood to be calculated in $|z\rangle_{r}$. The state $|z\rangle_{r}$ is called the squeezed state if the uncertainty at least in one of the observables $X$ or $P$ is less than $1/2$ \cite{MOScully97}.

\section{Resolution of unity and the Stieltjes moment problem}
\label{SecRes}

Since the first $r$ Fock states are absent from $|z\rangle_{r}$, the appropriate unity operator relevant for this situation is \cite{JMSixdeniers01, DPopov02}:
\begin{equation}\label{eq18}
I_{r} = \sum_{n=r}^{\infty} |n\rangle\langle n| = \sum_{n=0}^{\infty} |n+r\rangle\langle n+r|,
\end{equation}
which is an infinite sum of orthogonal projection operators, and the resolution of unity for the states $|z\rangle_{r}$ will read
\begin{equation}\label{eq19}
\int_{\mathbb{C}} d^2 z \, |z\rangle_{r} \,W_{r}(|z|^2)\,\,_{r}\langle z| = I_{r} = \sum_{n=0}^{\infty} |n+r\rangle\langle n+r|.
\end{equation}
Eq.(\ref{eq19}) is equivalent to an infinite set of integral conditions for a sought for positive function $W_{r}(|z|^2)$ that are obtained by performing the angular integration over $\theta$ ($z = |z| e^{i \theta}$):
\begin{equation}\label{eq20}
\int_{0}^{\infty} x^n \left[\pi \frac{W_{r}(x)}{\mathcal{N}_{r}(x)}\right] dx = \rho_{r}(n), \quad n=0, 1, \ldots , \quad x = |z|^2.
\end{equation}
This is the Stieltjes moment problem \cite{BSimon98, NIAkhiezer65} for $\widetilde{W}_{r}(x) = \pi W_{r}(x)/\mathcal{N}_{r}(x)$. The positivity requirement for $\widetilde{W}_{r}(x)$ can be always satisfied as the moment $\rho_{r}(n)$, see Eq.(\ref{eq15}), is a product of factorials and Eq.(\ref{eq20}) can be rewritten as a Mellin transform \cite{IASneddon72} of $\widetilde{W}_{r}(x)$ with $n = s-1$ (complex $s$) as follows:
\begin{eqnarray}\label{eq21}
\int_{0}^{\infty} x^{s-1} \widetilde{W}_{r}(x) dx &=& \mathcal{M}\,[\widetilde{W}_{r}(x); s] = \left[\prod_{k=0}^{r-1} \Gamma(s+k) \right]^2 \Gamma(s+r) \\[0.7\baselineskip]
&=& \rho_{r}(s-1). \label{eq21a}
\end{eqnarray}
The formal solution of Eq.(\ref{eq21a}) is $\widetilde{W}_{r}(x) = \mathcal{M}^{-1}[\rho_{r}(s-1); x]$, or equivalently
\begin{equation}\label{eq22}
\widetilde{W}_{r}(x) = \mathcal{M}^{-1} \left[\left(\prod_{k=0}^{r-1} \Gamma(s+k)\right)^2 \Gamma(s+r); x\right], \quad x\geq 0.
\end{equation}
In the above formulae $\mathcal{M}$ and $\mathcal{M}^{-1}$ denote the Mellin and inverse Mellin transforms respectively \cite{IASneddon72}. The positivity of $\widetilde{W}_{r}(x)$ follows from solving Eq.(\ref{eq21}) by the Mellin convolution \cite{OIMarichev83} which by itself conserves the positivity, see Refs. \cite{JRKlauder01, JMSixdeniers99, KAPenson10} for an elaboration of this property. 

\begin{figure}
\includegraphics[scale=0.6]{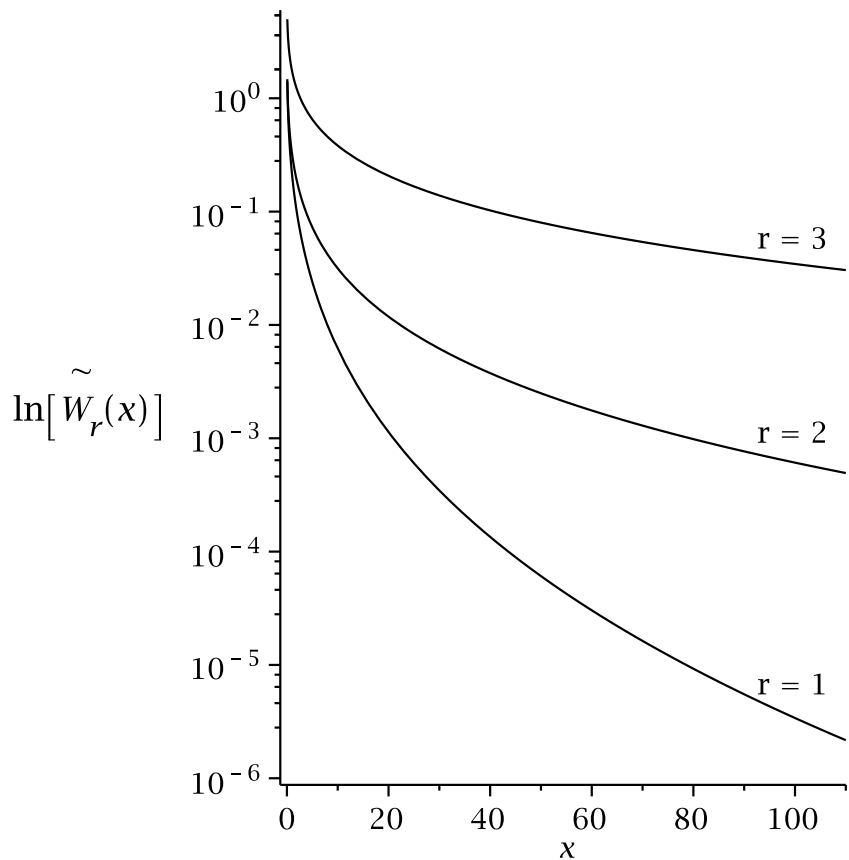}
\caption{\label{fig1} Plot of the logarithm of weight functions $\ln\left[\widetilde{W}_{r}(x)\right]$, see Eq.(\ref{eq23}), for $r = 1, 2$ and $3$, as a function of $x = |z|^2$. It is seen that for $x\geq 0$ the functions of Eq.(\ref{eq23}) are positive and are singular at $x=0$.}
\end{figure}

We can offer, for any $r$, explicit representations of $\widetilde{W}_{r}(x)$ in terms of Meijer's G functions $G(\ldots)$ \cite{APPrudniko98v} which, via Eqs.(\ref{eq21}) and (\ref{eq22}), are given by \cite{APPrudniko98v, OIMarichev83}:
\begin{equation}\label{eq23}
\widetilde{W}_{r}(x) = G([\,[\,\,\,], [\,\,\,]\,], [\,[\underbrace{0, 0, 1, 1, \ldots, r-1, r-1}_{2 r \, entries}, r], [\,\,\,]\,], x).
\end{equation}
We use the following transparent notation for Meijer's G function borrowed from computer-algebra systems \cite{przypis2}:
\begin{eqnarray}\nonumber
G([[list \,\, &of& \,\, l_{\alpha} \,\,parameters\,\, \alpha_{j} ], [list \,\, of\,\, l_{\gamma} \,\,parameters\,\, \gamma_{j}]], \\[0.7\baselineskip] \nonumber
&&[[list\,\, of \,\, l_{\beta} \,\, parameters\,\, \beta_{j}], [list\,\, of \,\,l_{\delta} \,\,parameters\,\, \delta_{j} ]], x) = \\[0.7\baselineskip] \label{eq23a}
&=& \mathcal{M}^{-1} \left[\frac{\prod_{j = 1}^{l_{\alpha}} \Gamma(1-\alpha_{j} - s) \, \prod_{j = 1}^{l_{\beta}} \Gamma(\beta_{j} + s)}{\prod_{j = 1}^{l_{\gamma}} \Gamma(\gamma_{j} + s) \,\prod_{j = 1}^{l_{\delta}} \Gamma(1 - \delta_{j} - s)} ; x \right].
\end{eqnarray}
Note that the parameter list in the third bracket in Eq.(\ref{eq23}) has $2 r + 1$ elements. Although apparently none of the  functions of Eq.(\ref{eq23}) can be represented by the known special functions we have at our disposal a rather complete knowledge about their characteristics \cite{APPrudniko98v, MDSpringer79}. We present on Fig. \ref{fig1} the plot of $\widetilde{W}_{r}(x)$ for $r = 1, 2, 3$ . Note that all $\widetilde{W}_{r}(x)$ are singular at $x=0$ for all $r = 1, 2, \ldots$. The solutions specified by Eq.(\ref{eq23}) are not unique, see Appendix ~A. 

We conclude that the states $|z\rangle_{r}$ are the coherent states in the sense of Klauder. They are normalized, see Eq.(\ref{eq6}), continuous with label and satisfy the resolution of unity with a positive weight function $\widetilde{W}_{r}(x)$. 

\section{Physical and statistical properties of coherent states $|z\rangle_{r}$}
\label{SecFig}

In this paragraph we shall present in detail the quantities enumerated in Eqs. (\ref{eq16})-(\ref{eq17c}) calculated for different values of $r$.

\begin{figure}
\includegraphics[scale=0.6]{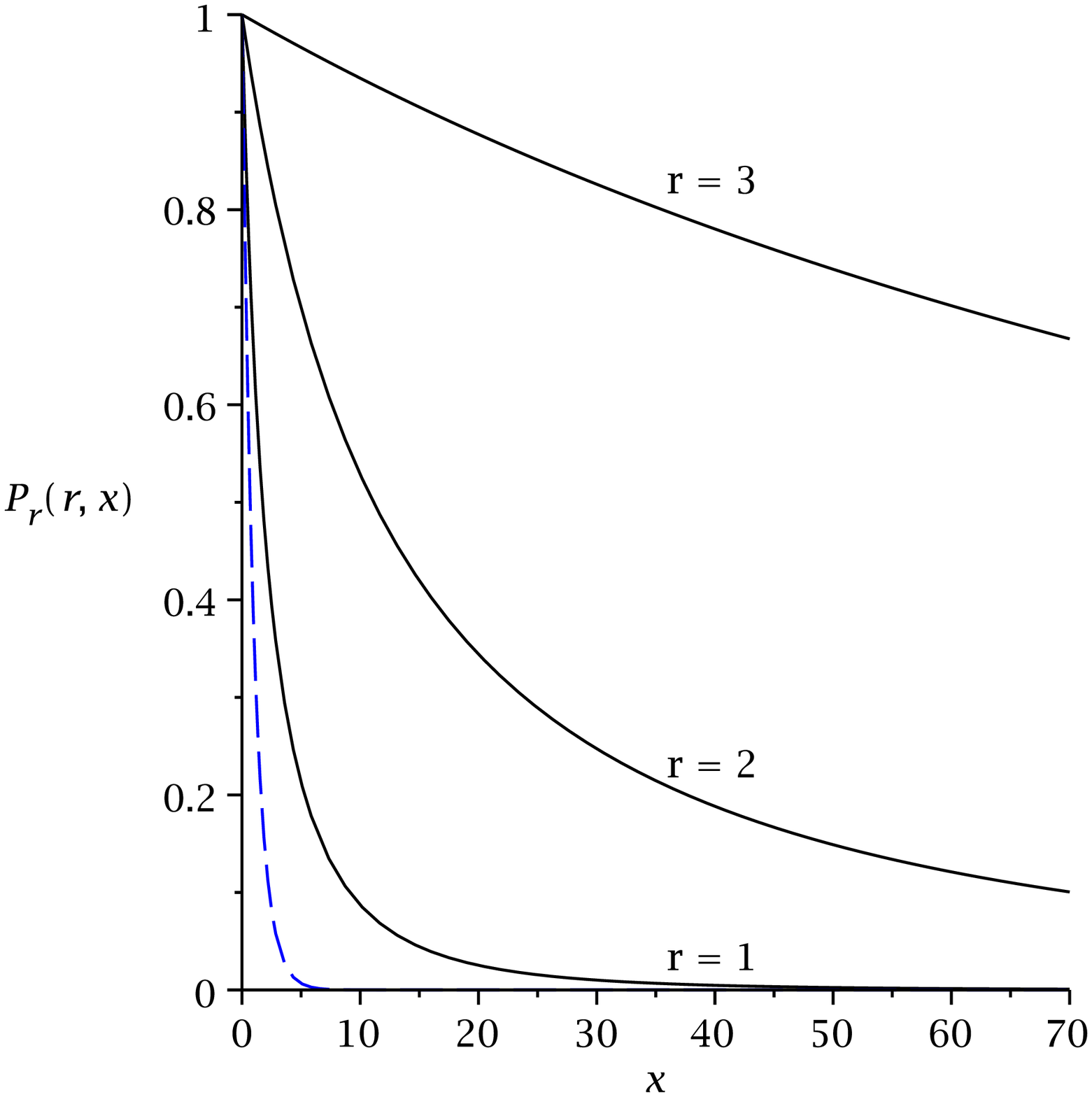}
\caption{\label{fig2} Plot of the probability $P_{r}(r, x)$ of finding the lowest allowed state $|r\rangle$ in the CS $|z\rangle_{r}$ for $r = 1, 2$ and $3$, as a function of $x = |z|^2$, see Eq.(\ref{eq16}). The dashed line is the probability of finding the state $|0\rangle$ in the standard CS, i.e. $e^{-x}$.}
\end{figure}

\begin{figure}
\includegraphics[scale=0.6]{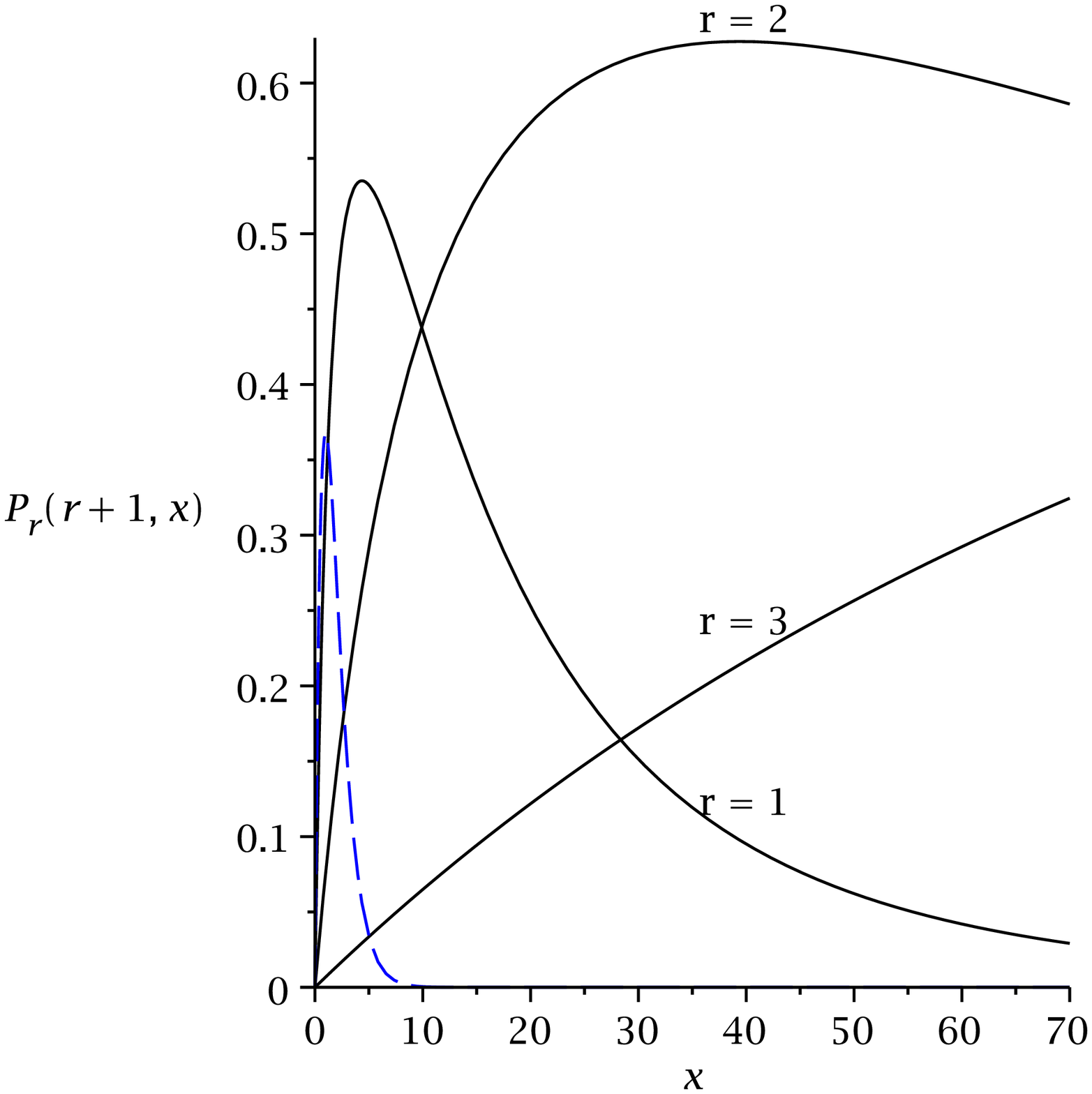}
\caption{\label{fig3} Plot of the probability $P_{r}(r+1, x)$ of finding the first allowed excited state $|r+1\rangle$ in the CS $|z\rangle_{r}$ for $r = 1, 2$ and $3$, as a function of $x = |z|^2$, see Eq.(\ref{eq16}).  All the curves display one maximum moving away from $x=0$ for increasing $r$. The dashed line is the probability of finding the state $|1\rangle$ in the standard CS, i.e. $x e^{-x}$.}
\end{figure}

The probability $P_{r}(k, x)$, defined in Eq.(\ref{eq16}), is presented in Figs. \ref{fig2} and \ref{fig3} as a function of $x$ for $r = 1, 2, 3$, for $k=r$ and $k=r+1$, along with the Poisson distribution originating from the standard CS. 

\begin{figure}
\includegraphics[scale=0.6]{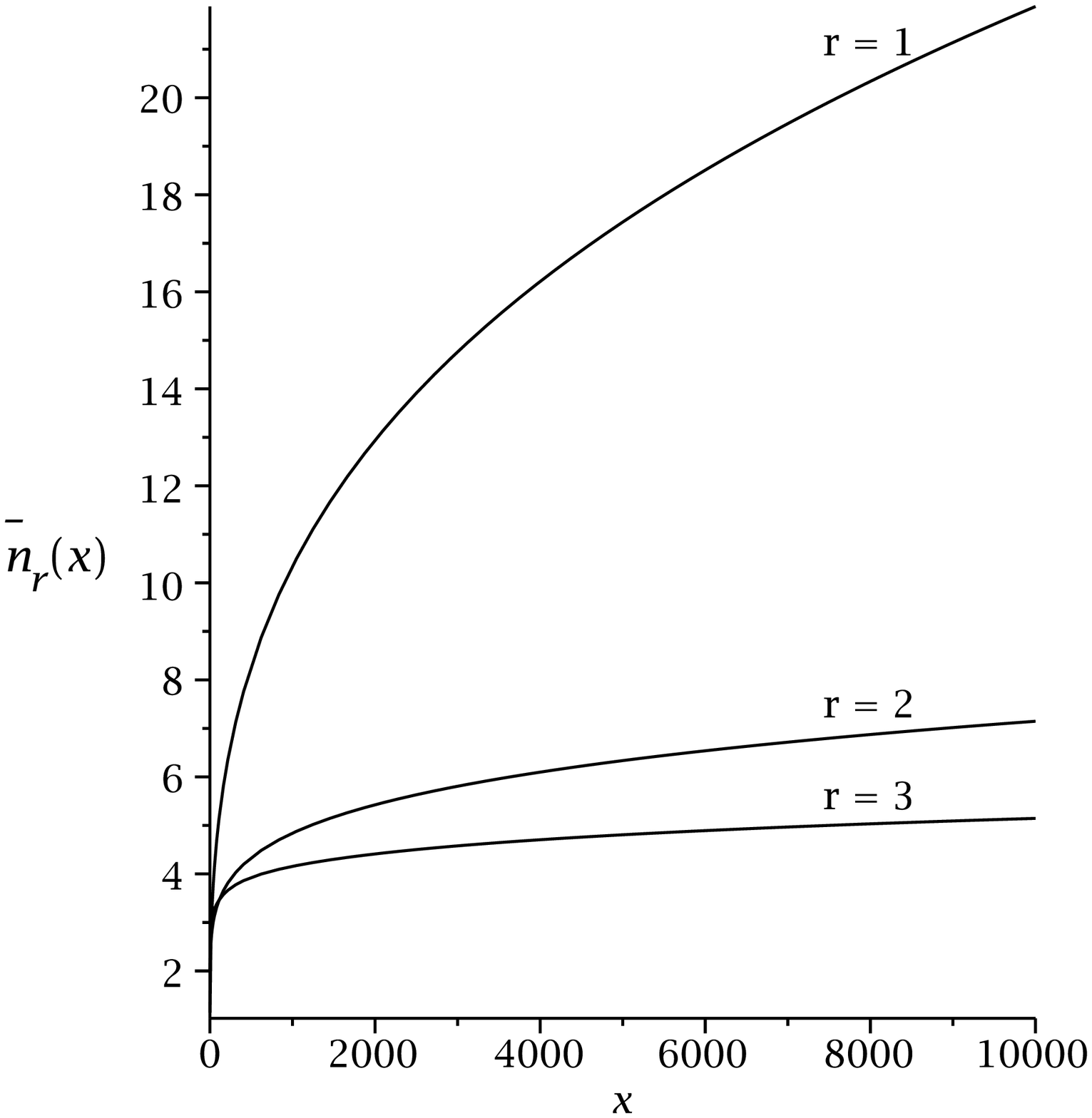}
\caption{\label{fig4} Plot of the average number $\bar{n}_{r}(x)$ of photons in the state $|z\rangle_{r}$, see Eq.(\ref{eq16a}), for $r = 1, 2$ and $3$, as a function of $x = |z|^2$.}
\end{figure}

The average numbers $\bar{n}_{r}(x)$ of photons in the state $|z\rangle_{r}$ are presented in Fig. \ref{fig4}. The dependence of $\bar{n}_{r}(x)$ for fixed $x$ as a function of $r$  indicates increasing deviation downwords from the straight line $n(x) = x$, characterising the standard CS. This behaviour clearly differs from the results describing the photon-added states considered in \cite{GSAgarwal91}, compare Fig.2 of this reference.

\begin{figure}
\includegraphics[scale=0.6]{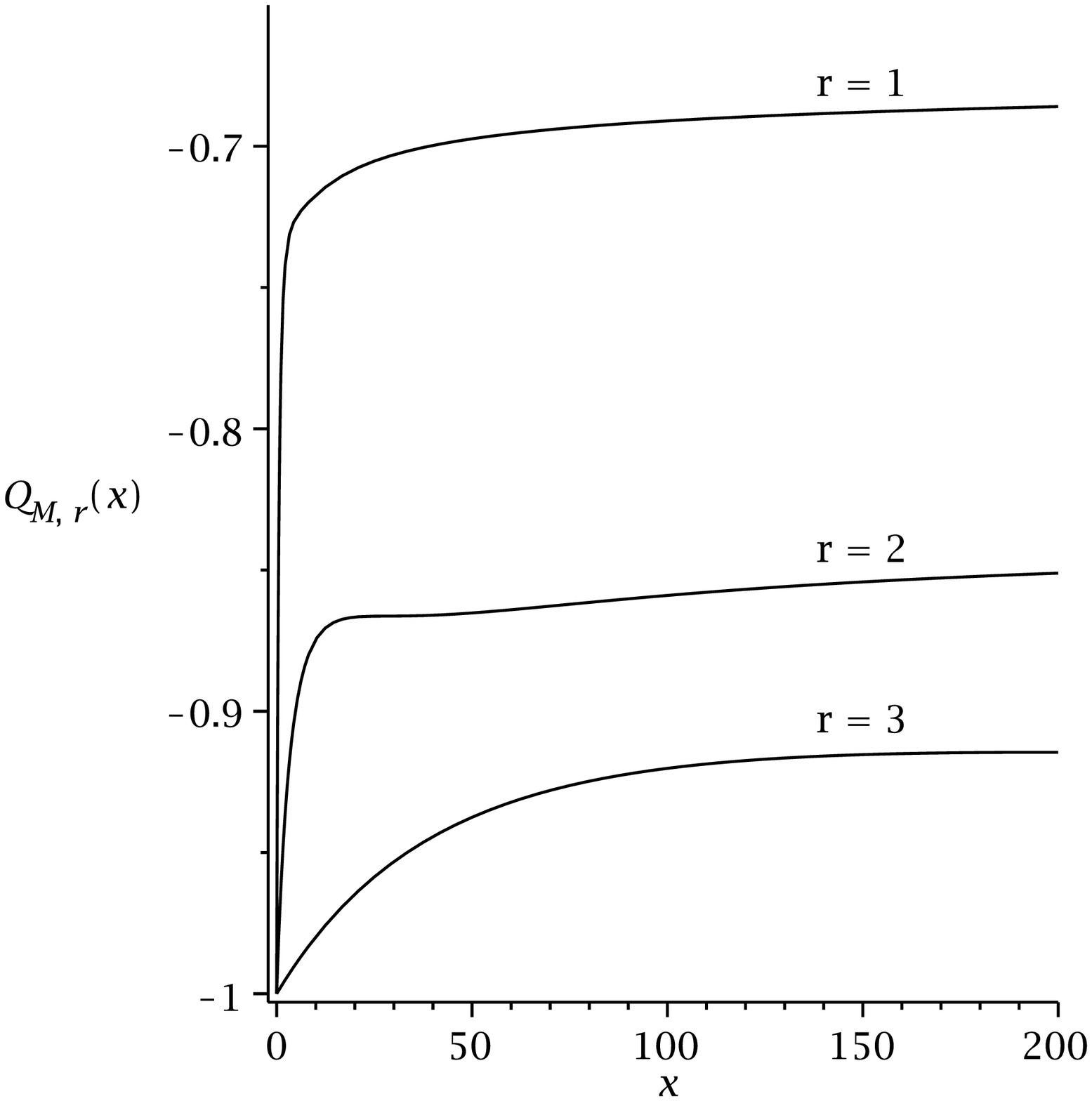}
\caption{\label{fig5} Plot of the Mandel parameter $Q_{M, r}(x)$ for $r=1, 2$ and $3$, as a function of $x=|z|^2$, see Eq.(\ref{eq17}). The states $|z\rangle_{r}$ are sub-Poissonian, because $Q_{M, r}<0$ for all $x>0$. Since for fixed $x$, $|Q_{M, r+1}| > |Q_{M, r}|$ the sub-Poissonian character is increasing with increasing $r$.}
\end{figure}

The Mandel parameter $Q_{M, r}(x)$ is presented in Fig. \ref{fig5}. For all $r$ the non-Poissonian character of the states $|z\rangle_{r}$ is explicit and is increasing with increasing $r$.

\begin{figure}
\includegraphics[scale=0.6]{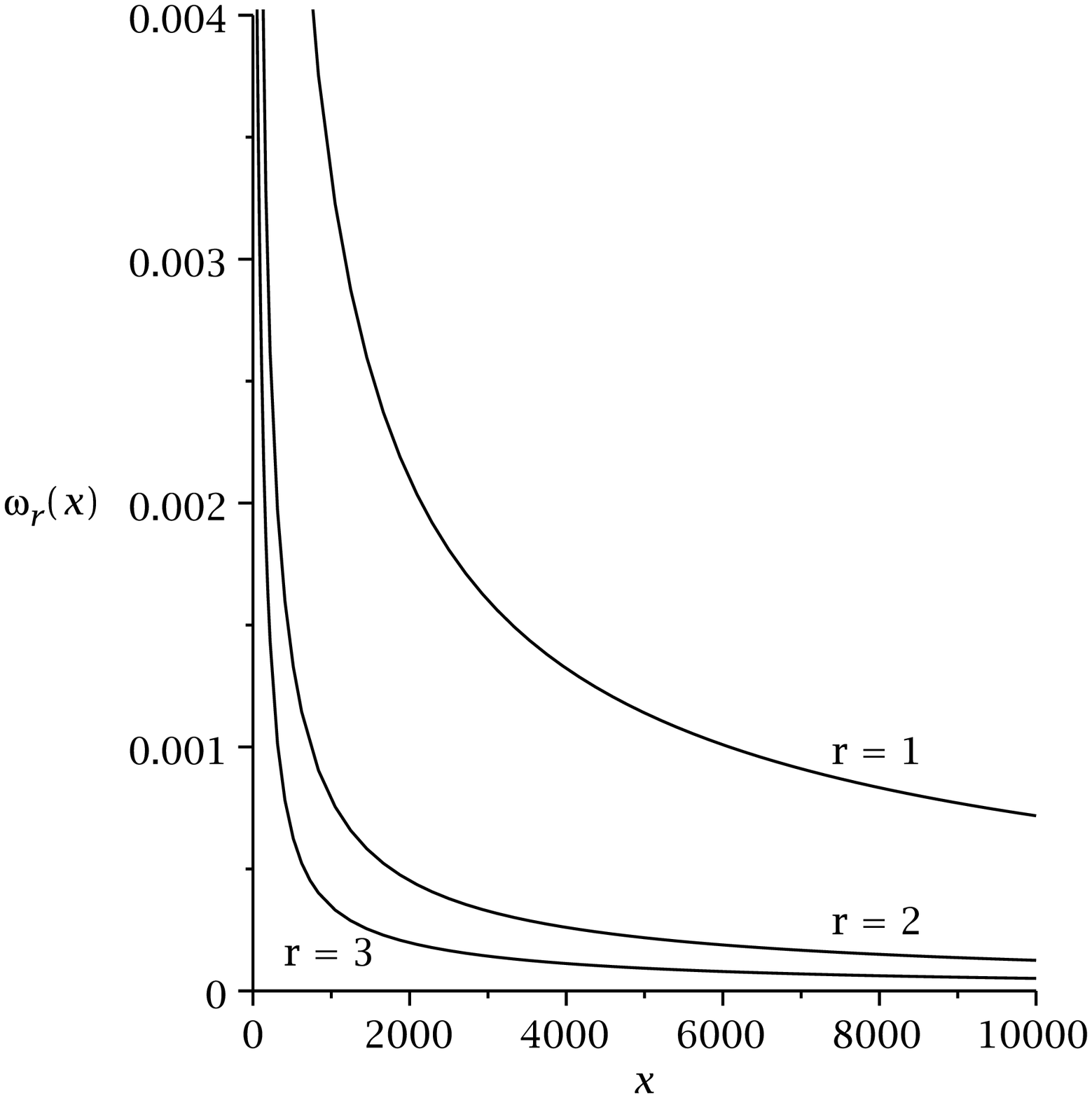}
\caption{\label{fig6} Plot of the metric factor $\omega_{r}(x)$ for $r = 1, 2$ and $3$, as a function of $x=|z|^2$, see Eq.(\ref{eq17a}).}
\end{figure}

In Fig. \ref{fig6} we present the plot of the metric factor $\omega_{r}(x)$ for $r = 1, 2, 3$. We observe the increasing deviation of $\omega_{r}(x)$ with increasing $r$ from the flat geometry of standard CS characterized by $\omega(x) = 1$ \cite{JRKlauder01}. 

\begin{figure}
\includegraphics[scale=0.6]{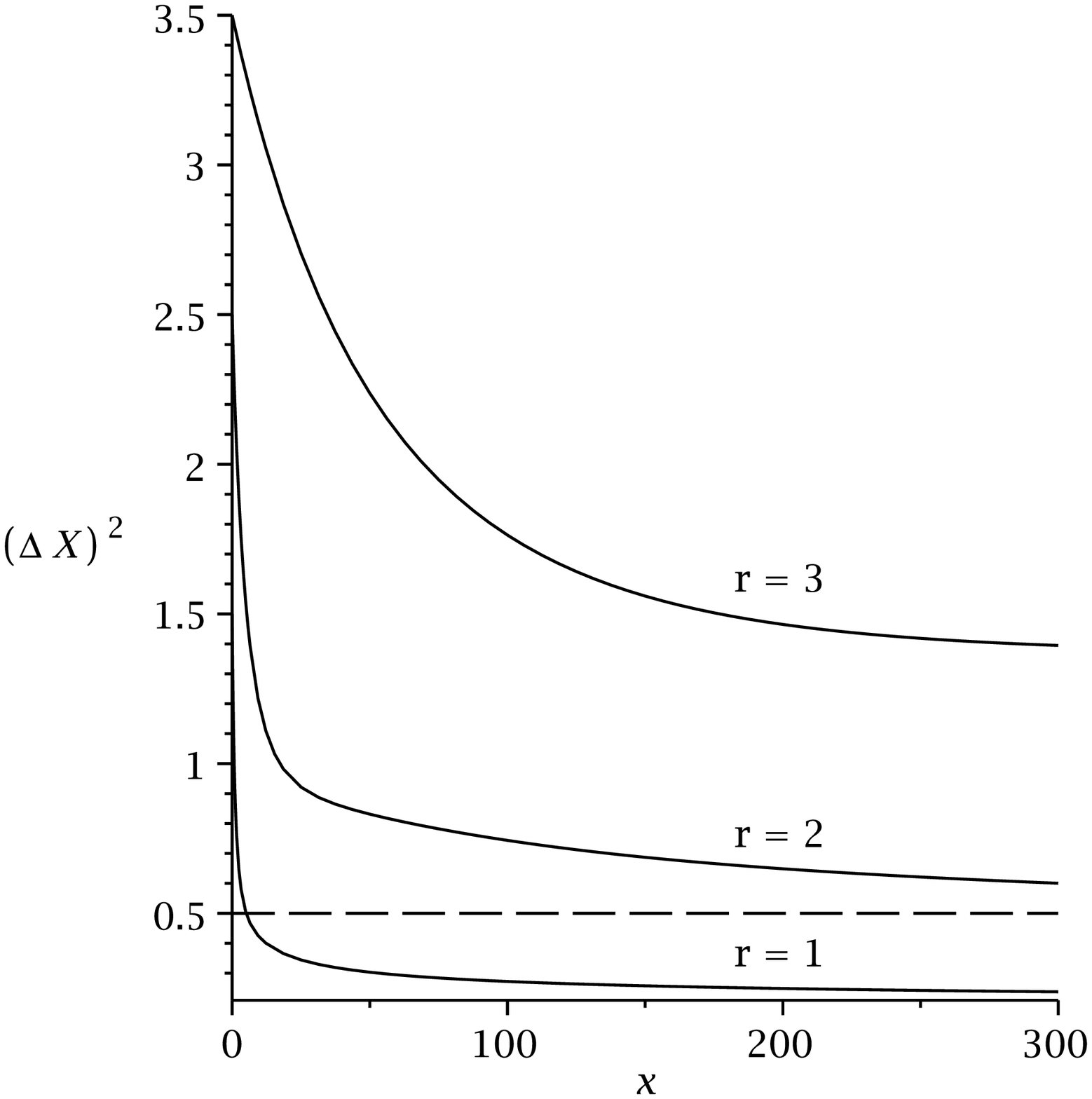}
\caption{\label{fig7} Plot of $(\Delta X)^2 = \overline{X^2} - \bar{X}^2$ as a function of $x = |z|^2$, for $r = 1, 2$ and $3$, see Eq.(\ref{eq17b}). The averages are calculated in the states $|z\rangle_r$.}
\end{figure}

\begin{figure}
\includegraphics[scale=1]{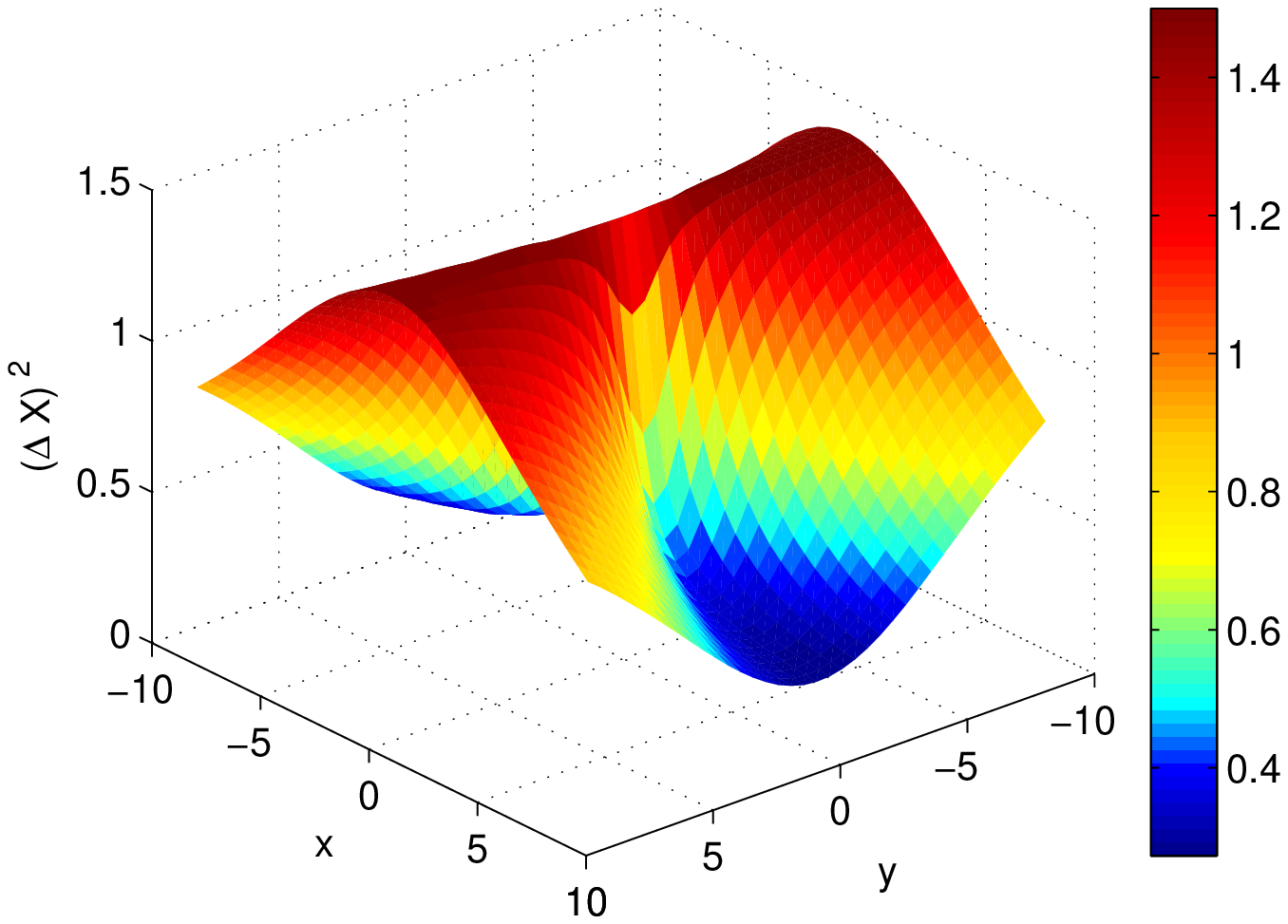}
\caption{\label{fig8} Plot of $(\Delta X)^2 = \overline{X^2} - \bar{X}^2$ for $|z\rangle_{1}$ as a function of $z = x + i\, y$.}
\end{figure}

\begin{figure}
\includegraphics[scale=1]{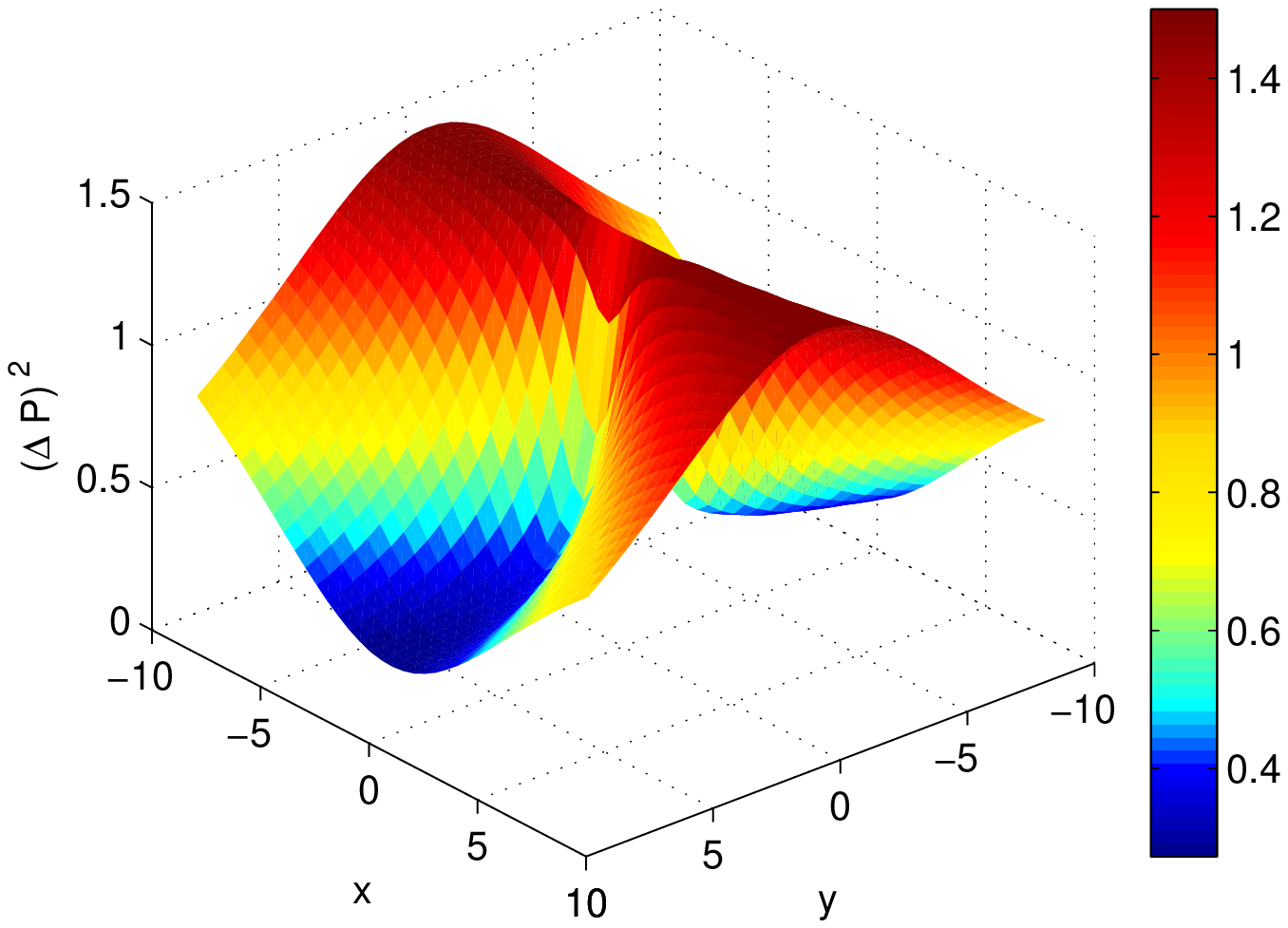}
\caption{\label{fig9} Plot of $(\Delta P)^2 = \overline{P^2} - \bar{P}^2$ for $|z\rangle_{1}$ as a function of $z = x + i\, y$. }
\end{figure}

In Fig. \ref{fig7} we present the uncertainty $(\Delta X)^2$ for several values of $r$. For sufficiently large $x$ for every $r$ $(\Delta X)^2$ becomes smaller $1/2$ (i.e. we observe squeezing in $x$), although in Fig. \ref{fig7} it is only visible for $r=1$ case. In Figs. \ref{fig8} and \ref{fig9} we present the squeezing in the complex plane for $X$ and $P$ quadratures respectively. As a guide for eye the squeezed regions are colored in blue. The Heisenberg uncertainty relation $\Delta X \cdot \Delta P \geq 1/2$ is evidently satisfied for all values of $r$. It can be already seen in Figs.~\ref{fig8} and~\ref{fig9}, for $r=1$, by reading off the appropriate values for a given $x$. A similar behaviour persists for other values of $r$ and corresponding schemes are not reproduced here.

\section{Discussion and Conclusion}
\label{SecDis}

We have initiated in this work a method of generating collective quantum states of the Fock space via the action of certain non-unitary analogues of conventional boson displacement operator. The prescription involves first finding eigenfunctions $E_{r}(x)$ of simple classical differential operators. Then the non-unitary analogue of displacement operator is obtained by putting the boson creation operator as the argument in those eigenfunctions $E_{r}(z\,a^{\dag})$ with complex $z$. This procedure allows in a systematic way the construction of quantum states $|z\rangle_{r}$ via action $E_{r}(z a^{\dag})|0\rangle$, see Eq.(\ref{eq5a}) which is the lynchpin of this method. In other words we depart from the exponential as a vehicle to create the states and look for other, related families of functions (here of hypergeometric type) for that purpose. In this paper we have limited ourselves to an operator of the form $x^r d^{r+1}$ which quite naturally produces eigenfunctions of $(a^{\dag})^r\, a^{r+1}$ with the eigenvalue $z$. The fact that the quantum CS so obtained almost automatically fulfil the criteria of Klauder, is quite satisfying and may give rise to many possible extensions.

Since our state $|z\rangle_{r}$ is reminiscent of photon added-state of Agarwal et al, it is natural to compare the properties of $|z\rangle_{r}$ with those of states of Ref. \cite{GSAgarwal91}. Since their Fock space expansions are different (compare our Eq.(\ref{eq7}) with Eq.(2.6) of Ref. \cite{GSAgarwal91}), one expects that their physical and statistical properties differ too. In fact this is the case and it appears that there is no way to relate these two families of coherent states. However, there are some common features: both states are squeezed, are sub-Poissonian in nature and possess the resolution of unity (unique for states of Ref. \cite{GSAgarwal91}, see Ref. \cite{JMSixdeniers01}, and non-unique for states considered in the present work). It is interesting to observe that the eigenproperties of these families can be obtained exactly: the relevant operator in our case is quite natural as it is given by the boson form of differential operator whose eigenfunctions we consider. In the other case the appropriate operator is quite involved and has been obtained some time after the states have been proposed \cite{SSivakumar99}. 

\section{Acknowledgements}

The authors acknowledge support from Agence Nationale de la Recherche (Paris, France) under Program No. ANR-80-BLAN-0243-2 and from PAN/CNRS Project PICS No.4339 (2008-2010).

\appendix
\section{}

A special feature of the solutions of the Stieltjes moment problem, Eq.(\ref{eq20}) is the fact that the $\widetilde{W}_{r}(x)$ defined by Eq.(\ref{eq22}) and (\ref{eq23}) and visualized in Fig. \ref{fig1}, are not unique solutions for all $r=1, 2, \ldots$. It means that there exist other \textit{positive} solutions $V_{r}(x)\neq\widetilde{W}_{r}(x)$ such that $\int_{0}^{\infty} x^n V_{r}(x) dx = \rho_{r}(n)$, $n = 0, 1, \ldots$. In order to prove the non-unique character of solutions of Eq.(\ref{eq20}) we apply the following sufficient condition for non-uniqueness of the Stieltjes moment problem; it involves both the moments $\rho_{r}(n)$ and the solution $\widetilde{W}_{r}(x)$ \cite{KAPenson10}: If
\begin{itemize}
\item[(a) ] $S_{r} = \sum_{n=0}^{\infty} [\rho_{r}(n)]^{-1/2n} < \infty$, and

\item[(b) ] the function $\psi_{r}(x) = -\ln\left[\widetilde{W}_{r}(\exp(x))\right]$ is convex for $x > 0$, 
\end{itemize}
then $\widetilde{W}_{r}(x)$ in non-unique \cite{KAPenson10, APakes01, APakes07, AGut02}.\\
\noindent
Concerning (a): we establish, via logarithmic test of convergence \cite{APPrudniko98v}, that $S_{r} < \infty$ for all $r = 1, 2, \ldots$ . For (b) it suffices to prove that for fixed $r$ the condition $[\psi_{r}(x)]''\geq 0$ is fulfilled. In the general case it is difficult to show it analytically, but it can be done relatively easy with the help of computer algebra systems \cite{przypis2}.

The actual construction of such non-unique solutions, i.e. finding functions $V_{r}(x)$ referred to above, is still an open and challenging problem, see \cite{KAPenson10} and reference therein.

\end{document}